\documentclass[twocolumn]{aastex631}
\usepackage{xcolor}
\usepackage[hyphens]{url}

\usepackage{xcolor}

\newcommand*{\Chandra}{Chandra}
\defcitealias{Mantz2016}{M16}
\newcommand*{\Msixteen}{\citetalias{Mantz2016}}

\begin{document}

\title{SPT-CL J0417$-$4748: A Deep \Chandra{} Study of a Relaxed Galaxy Cluster Without Central Star Formation}

\author[0000-0003-3521-3631]{Taweewat Somboonpanyakul}
\affiliation{Department of Physics, Faculty of Science, Chulalongkorn University, 254 Phayathai Road, Pathumwan, Bangkok 10330, Thailand}
\affiliation{Kavli Institute for Particle Astrophysics and Cosmology, Stanford University, 452 Lomita Mall, Stanford, CA 94305, USA}

\author[0000-0002-8031-1217]{Adam B. Mantz}
\affiliation{Kavli Institute for Particle Astrophysics and Cosmology, Stanford University, 452 Lomita Mall, Stanford, CA 94305, USA}

\author[0000-0003-0667-5941]{Steven W. Allen}
\affiliation{Kavli Institute for Particle Astrophysics and Cosmology, Stanford University, 452 Lomita Mall, Stanford, CA 94305, USA}
\affiliation{Department of Physics, Stanford University, 382 Via Pueblo Mall, Stanford, CA 94305, USA}
\affiliation{SLAC National Accelerator Laboratory, 2575 Sand Hill Road, Menlo Park, CA 94025, USA}

\author[0000-0001-7179-6198]{Anthony M. Flores}
\affiliation{Kavli Institute for Particle Astrophysics and Cosmology, Stanford University, 452 Lomita Mall, Stanford, CA 94305, USA}
\affiliation{Department of Physics, Stanford University, 382 Via Pueblo Mall, Stanford, CA 94305, USA}
\affiliation{Department of Physics and Astronomy, Rutgers University 136 Frelinghuysen Rd Piscataway, NJ 08854, USA}

\author[0000-0003-2985-9962]{R. Glenn Morris}
\affiliation{Kavli Institute for Particle Astrophysics and Cosmology, Stanford University, 452 Lomita Mall, Stanford, CA 94305, USA}
\affiliation{Department of Physics, Stanford University, 382 Via Pueblo Mall, Stanford, CA 94305, USA}
\affiliation{SLAC National Accelerator Laboratory, 2575 Sand Hill Road, Menlo Park, CA 94025, USA}

\author[0000-0002-2776-978X]{Haley R. Stueber}
\affiliation{Kavli Institute for Particle Astrophysics and Cosmology, Stanford University, 452 Lomita Mall, Stanford, CA 94305, USA}
\affiliation{Department of Physics, Stanford University, 382 Via Pueblo Mall, Stanford, CA 94305, USA}

\author[0000-0001-7665-5079]{Lindsey E. Bleem}
\affiliation{High-Energy Physics Division, Argonne National Laboratory, 9700 South Cass Avenue., Lemont, IL, 60439, USA}

\author[0000-0003-4175-571X]{Benjamin Floyd}
\affiliation{Institute of Cosmology \& Gravitation, University of Portsmouth, Dennis Sciama Building, Portsmouth, PO1 3FX, UK}

\author[0000-0001-7271-7340]{Julie Hlavacek-Larrondo}
\affiliation{D\'{e}partement de Physique, Universit\'{e} de Montr\'{e}al, Succ. Centre-Ville, Montr\'{e}al, Qu\'{e}bec, H3C 3J7, Canada}

\author[0000-0001-6505-0293]{Keunho J. Kim}
\affiliation{IPAC, California Institute of Technology, 1200 E. California Boulevard, Pasadena, CA 91125, USA}

% \collaboration{10}{(SPT collaboration)}

\email{taweewat.s@chula.ac.th}



\begin{abstract}
\noindent 
We present an in-depth \Chandra{} X-ray analysis of the galaxy cluster SPT-CL J0417$-$4748 (hereafter SPT\,J0417), at $z = 0.58$, with a focus on its thermodynamic properties and the apparent absence of central star formation. Utilizing a total \Chandra{} exposure of 103 ks, we find that the large-scale X-ray morphology is consistent with a dynamically relaxed, cool-core system. The intracluster medium (ICM) shows a central density of \(0.08\pm0.01\) \(\rm{cm^{-3}}\), a central pseudo-entropy of \(26^{+6}_{-5}\,\rm{keV\,cm^{2}}\) and a central cooling time of $515^{+96}_{-75}$ Myr, values typical of massive cool-core clusters. Despite these conditions, no evidence of recent or ongoing star formation is detected in the brightest cluster galaxy (BCG). Spectral energy distribution (SED) fitting of DES photometry indicates that the bulk of the stellar population formed at $z\sim1.25$, with no significant star formation over the past $\sim$3 Gyr, while optical spectra from Magellan show no [O II] emission. Complementary ASKAP radio and Spitzer infrared data indicate a lack of strong current AGN activity in the BCG. SPT\,J0417 exemplifies massive, relaxed, cool-core clusters in which cooling and star formation appear almost completely quenched, providing valuable insights into how AGN feedback regulates the long-term thermal balance of the intracluster medium.

% we measured the cooling time, pseudo-entropy, metallicity, and temperature structure of the cluster. While SPT\,J0417 exhibits a strong cool-core profile with a central cooling time of $515^{+96}_{-75}$ Myr and an X-ray luminosity (\(1.08\pm0.03 \times 10^{45}\,\rm{erg\,s^{-1}}\)), consistent with other massive cool-core clusters, the bright central galaxy (BCG) shows no evidence of ongoing star formation, as indicated by the absence of [O II] emission in the optical spectra obtained using the Magellan telescope. Complementary radio and infrared data from ASKAP and Spitzer, respectively, further indicate a lack of radio-luminous AGN activity in the BCG. 
% \todo{The cluster's slightly elevated central pseudo-entropy and relatively flat central density profile of the intracluster gas suggest that subtle perturbations, potentially associated with past dynamical activity, may have suppressed cooling.} These findings place SPT\,J0417 in a rare class of massive, cool-core clusters where star formation is effectively quenched, providing valuable insights into AGN feedback and the regulation of cooling in the intracluster medium.
\end{abstract}

\keywords{Galaxy clusters (584)}

%% and \citet commands to identify citations. The citations are

\section{Introduction} \label{sec:intro}
Early X-ray observations of galaxy clusters revealed a hot (\(\sim10^7\) K), X-ray-emitting plasma known as the intracluster medium (ICM), permeating the cluster environment~\citep{Fabian1977,White1991,Edge1992,Voit2005}. This plasma is the dominant baryonic matter component of galaxy clusters. Some systems, known as "cool-core" clusters, exhibit high and sharply peaked central ICM densities and relatively low temperatures~\citep{Allen2001,McNamara2007,Werner2014}. In the absence of a heating mechanism, the gas in cool cores would rapidly cool and flow inward, generating a ``cooling flow''~\citep{Sarazin1986,Fabian1994}. This process would deposit large amounts of cold gas onto the bright central galaxy (BCG), fueling significant star formation.

However, observations reveal much lower star formation rates than predicted, a discrepancy referred to as the ``cooling flow problem''~\citep{ODea2008,Donahue2015,McDonald2018}. Active galactic nucleus (AGN) feedback, driven by the energetic output of supermassive black holes (SMBHs) residing in cluster cores, is thought to play a central role in regulating radiative cooling by coupling to the surrounding intracluster medium and offsetting excessive energy losses~\citep{Fabian2012,McNamara2012}. This feedback manifests through radiation, wind, and/or relativistic jets~\citep{Rafferty2006,McNamara2007,Hlavacek-Larrondo2013}. 
%~\citep{Hlavacek-Larrondo2013,Hlavacek-Larrondo2015,McDonald2013,Birzan2017}.

Over the past few decades, observations have revealed a wide range of star formation rates (SFRs) in cool-core BCGs, from extreme cases such as the Phoenix Cluster with SFRs of \(\sim\!\!800~M_{\odot}~\rm{yr}^{-1}\), %to more moderate values of \(\sim\!\!120\) \(\rm{M_{\odot}~yr^{-1}}\) in Abell 1835, 
down to just a few \(\rm{M_{\odot}~yr^{-1}}\) in systems like Abell 478 and RX J1720.1+2638, and even lower rates (\(<\!1\) \(\rm{M_{\odot}~yr^{-1}}\)) in Abell 2029~\citep{Cavagnolo2009,McDonald2013,Fraser2014,McDonald2018}. These findings have led to several theoretical models exploring the relationships between SFR, ICM cooling, and AGN feedback.

One proposed scenario suggests that some massive cool-core clusters exhibit high SFRs (\(>\!\!100~M_{\odot}~\rm{yr}^{-1})\) owing to a saturation in AGN feedback heating, particularly in the most massive systems~\citep{McDonald2018,Calzadilla2022}. However, the existence of many massive cool-core clusters with little to no star formation in their BCGs indicates that not all such systems have reached this saturation threshold. This raises important questions regarding the physical conditions under which AGN feedback becomes inefficient at regulating cooling. Identifying more examples of massive cool-core clusters with a broad range of SFRs will provide further insight into the underlying mechanisms governing cooling and feedback, and their role in the broader context of galaxy cluster evolution.

SPT\,J0417 (\(z = 0.58\)) represents an intriguing case of a massive, cool-core cluster exhibiting minimal star formation activity in its BCG, indicating that AGN feedback may remain below the saturation threshold. Identified in the 2500 \(\rm{deg}^2\) South Pole Telescope-Sunyaev-Zel'dovich (SPT-SZ) cluster survey~\citep{Williamson2011,Bleem2015}, this cluster was observed with the \Chandra{} X-ray Observatory for 22 ks in 2013. \citet{McDonald2013,McDonald2016} classified this object as a typical cool-core system, with no evidence of star formation in the central galaxy. In this study, we present a deeper observation of SPT\,J0417 using combined \Chandra{} data totaling 110 ks in exposure time. These observations enable a more detailed exploration of the cluster core and provide an opportunity to investigate the potential mechanisms that could suppress star formation in the cluster's center.

The remainder of this paper is organized as follows. Section~\ref{sec:observation} outlines the \Chandra{} X-ray observations and data reduction, and presents supporting multiwavelength datasets, including infrared, radio, and optical spectroscopic observations of BCG. The primary X-ray analysis and results are detailed in Section~\ref{sec:results}, and a summary of the findings and their broader implications are presented in Section~\ref{sec:conclusion}. Throughout this work, we adopt \(H_0 = 70\) km\,s$^{-1}$Mpc$^{-1}$, \(\Omega_m = 0.3\), and \(\Omega_\Lambda = 0.7\), with all uncertainties quoted at the 68.3\% level, unless otherwise stated. 

\section{Data} \label{sec:observation}
\subsection{X-ray: \Chandra{}}
SPT\,J0417 was initially observed with the \Chandra{} ACIS-I detector for 18 ks (after cleaning) in 2013. A follow-up observation conducted in 2023 using the ACIS-S detector added 85 ks of exposure. A combined total of 103 ks of clean time was analyzed using CIAO v4.16 and CALDB v4.11.0. Within $r_{500}$ (990 kpc), the combined dataset contains an estimated 10,228 net counts in the 0.6 to 7.0 keV band, corresponding to a count rate of 0.0993 $\rm{count\,s^{-1}}$.

The procedures for reducing and cleaning the \Chandra{} data follow the guidelines provided in the \Chandra{} Analysis Guide\footnote{\url{https://cxc.harvard.edu/ciao/guides/acis_data.html}}, as described in~\citet{Mantz2015}. In addition, we applied a time-dependent correction to the ancillary response files used in spectral modeling, as motivated by long-term calibration analyses of Abell 1795 over the course of the mission~\citep{Mantz2025}.

\begin{figure*}[ht!]
\includegraphics[width=0.98\textwidth]{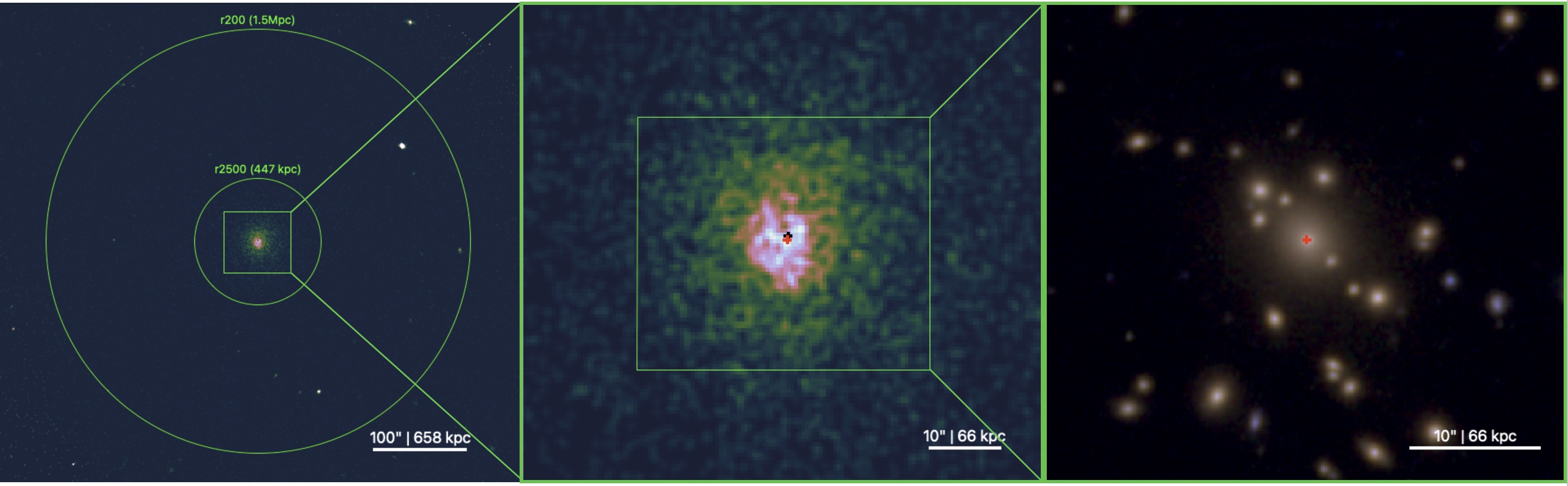}
\caption{The left panel presents a 0.6–7.0 keV \Chandra{} X-ray image of the cluster's central region, constructed from raw pixels of $0.492\times0.492$ $\rm{arcsec^{2}}$ and smoothed with a Gaussian filter of 2-pixel (0.984 arcsec) radius. The red crosses indicate the location of the BCG. The SPA center from the morphological analysis is within $0.5^{\prime\prime}$ of the BCG center. The right panel displays the Euclid image~\citep{Euclid2405.13491}, revealing a central red giant elliptical galaxy with other similarly red galaxies nearby. \label{fig:chandra}}
\end{figure*}

Figure~\ref{fig:chandra} displays the 0.6--7.0 keV \Chandra{} X-ray image of SPT\,J0417, smoothed with a Gaussian kernel of 2-pixel (0.984 arcsec) radius, to emphasize the cluster's central region. The green circles denote the characteristic radii $r_{2500}$ and $r_{500}$ derived from the best-fitting Navarro-Frenk-White (NFW) mass model~\citep{Navarro1997} (see Section~\ref{sec:spec_analysis}). The central panel presents a zoomed-in view of the X-ray image, highlighting a core morphology that is moderately peaked and exhibits slight asymmetry on scales of a few arcseconds. The right panel shows the Euclid image of the cluster core in  SPT\,J0417~\citep{Euclid2405.13491}, revealing a massive red elliptical galaxy identified as the BCG (marked by the red cross), surrounded by similarly red elliptical galaxies.

\begin{figure*}[ht!]
\centering
\includegraphics[width=0.98\textwidth]{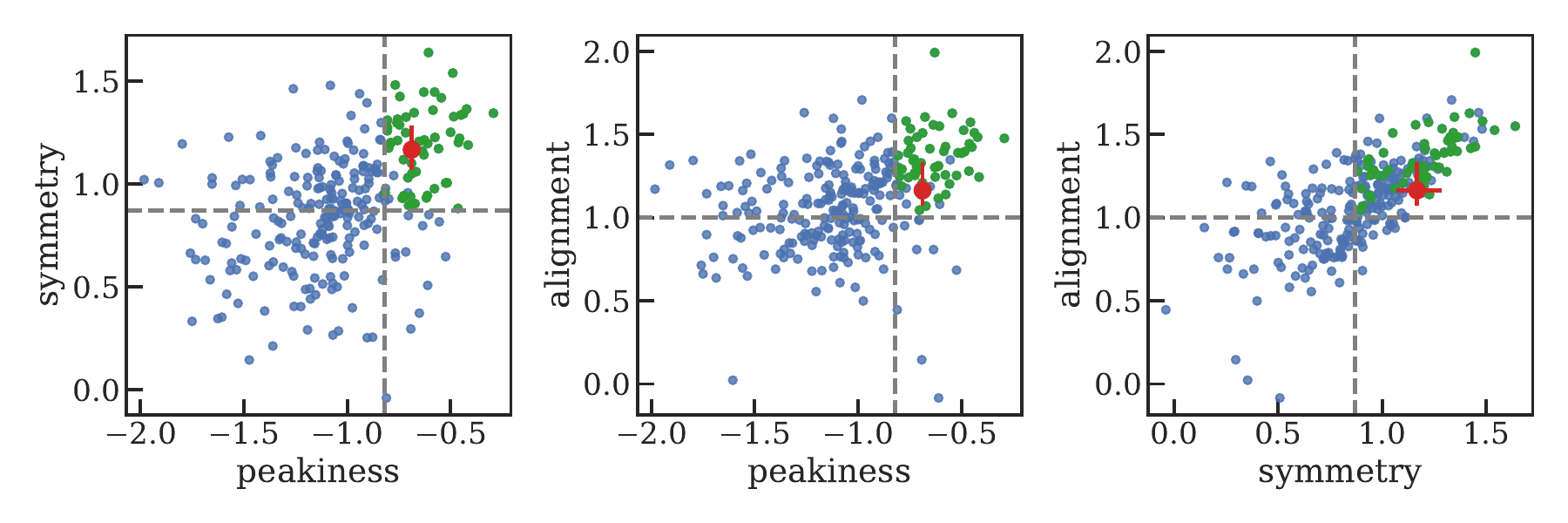}
\caption{Distribution of X-ray peakiness ($p$), symmetry ($s$), and alignment ($a$) for clusters from~\citet{Mantz2015}. The dashed lines indicate the threshold used to define the relaxed sample. Clusters meeting all three criteria are shown in green, while non-relaxed clusters are plotted in blue. The red point marks SPT\,J0417, which lies well within the relaxed region. 
\label{fig:spa}}
\end{figure*}

From the X-ray image, we characterized the X-ray morphology of the cluster using the symmetry–peakiness–alignment (SPA) metrics of~\citet{Mantz2015}, which quantify the sharpness of the surface brightness peak (\(p\)), as well as the symmetry and alignment of elliptical isophotes (\(s\) and \(a\)). The analysis yielded \(p = -0.688_{-0.007}^{+0.009}\), \(s = 1.17_{-0.10}^{+0.12}\), and \(a = 1.16_{-0.09}^{+0.17}\), firmly classifying the cluster as dynamically relaxed based on the relaxation thresholds of \(p > -0.82\), \(s > 0.87\), and \(a > 1.00\). In Figure~\ref{fig:spa}, this is illustrated by the red point (SPT\,J0417), which lies comfortably within the region of relaxed cluster (green points) in the SPA parameter space. We also find that the SPA center derived from the ICM morphology is offset by less than $0.5^{\prime\prime}$ from the BCG center, further supporting the classification of the system as dynamically relaxed. %**Add a note about agreement of X-ray and BCG centers (quote values), and how this provides further evidence that the cluster is relaxed. ***

\subsection{Radio: ASKAP}
\begin{figure*}[ht!]
\includegraphics[width=0.98\textwidth]{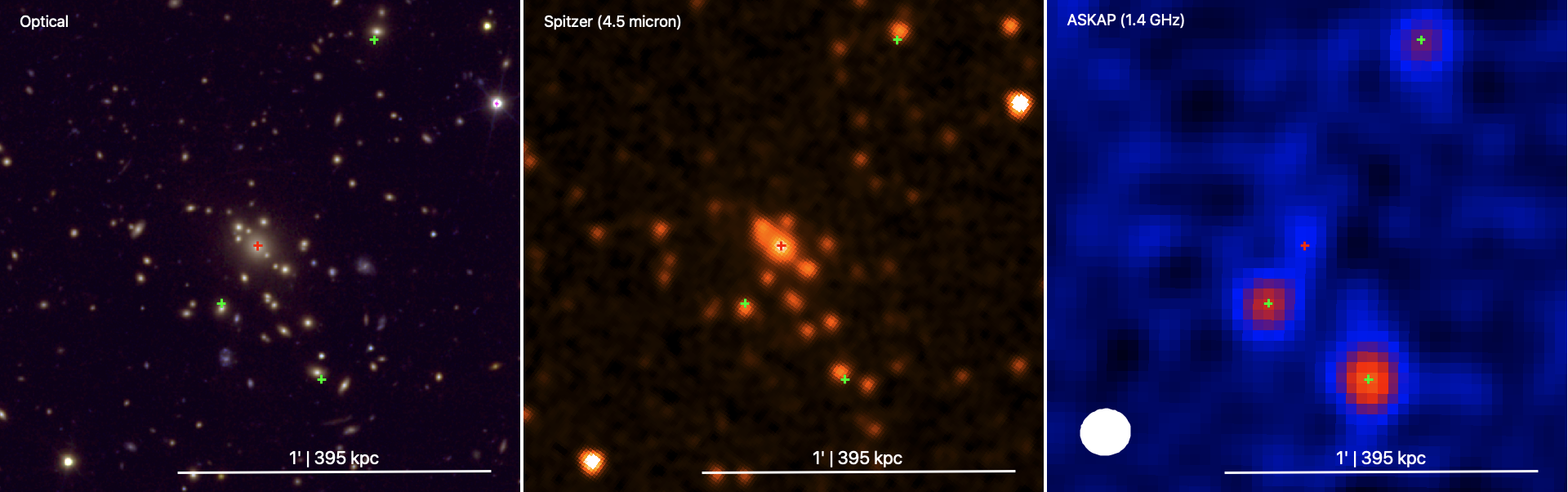}
\caption{The left panel displays the optical image of the cluster from Euclid~\citep{Euclid2405.13491} for reference, with red crosses indicating the position of the BCG and green crosses marking the locations of detected radio sources. The middle panel shows the 4.5\,$\mu$m infrared image obtained with \textit{Spitzer}, providing a view of the stellar and dust content. The right panel presents the 1.4\,GHz radio image from ASKAP with the beam size shown in the bottom left. Several radio sources are detected near the cluster center, but none is coincident with the BCG; instead, each is spatially associated with an optical counterpart confirmed to be a cluster member, making an association with the BCG unlikely. From these data, we derive a $3\sigma$ upper limit on the BCG radio luminosity of \(\sim\!1\times10^{40}~\rm{erg~s^{-1}}\).
\label{fig:radio_ir}}
\end{figure*}

For radio data, we used 1367 MHz observations from the Rapid ASKAP Continuum Survey (RACS)~\citep{Hale2021}. RACS offers an angular resolution of \(8^{\prime\prime}\) and provides uniform sensitivity (\(\sim0.3\) mJy rms) across a large area of the sky~\citep{Duchesne2023}. Three radio sources were detected by RACS at projected separations of approximately \(13^{\prime\prime}\), \(28^{\prime\prime}\), \(44^{\prime\prime}\) from the BCG~\citep{Duchesne2024}. The right panel of Figure~\ref{fig:radio_ir} shows the 1.4\,GHz ASKAP radio image, highlighting the three radio sources.

SPT\,J0417 is also included in the Dark Energy Survey (DES) Year 3 optically-selected ``redMaPPer" cluster catalog~\citep{DES2025}. Each of the three ASKAP radio sources is associated with a confirmed cluster member galaxy, with positional offsets of less than \(\sim\!1.6^{\prime\prime}\), well below the RACS beam size. The assigned photometric redshifts of the matched member galaxies are $0.58\pm0.02$ (East object), $0.60\pm0.02$ (West object) and $0.59\pm0.02$ (North object) with cluster membership probabilities of 0.993, 0.995 and 0.995, respectively~\citep{DES2025}. %These measurements suggest that all of these galaxies are cluster members and host the detected radio emissions.
% Spectroscopic redshift confirmation in both radio and optical is required to verify that these radio sources are indeed part of the cluster.
%from DES Y3
%membership probability (left): 0.99258363, photo_z = 0.58833474+/-0.02104005 (0.9")
%membership probability (right): 0.99467254, photo_z = 0.6013767+/-0.020016925 (1.6")
%membership probability (northern): 0.99467254, photo_z = 0.5919332+/-0.018282041 (1.6")

For the BCG, we report an upper limit on the radio luminosity of \(\sim\!1\times10^{40}~\rm{erg~s^{-1}}\), corresponding to three times the rms noise of the ASKAP map and following the approach of~\citet{Calzadilla2024}. This limit rules out the presence of particularly prominent relativistic radio jets or a radio-luminous AGN in the central galaxy. However, radio emission below the current sensitivity cannot be excluded, as deeper observations have revealed BCGs hosting radio sources with luminosities below $10^{39}\,\rm erg\,s^{-1}$~\citep[e.g.,][]{Veronica2025}. More generally, AGN feedback in BCGs is expected to be episodic, with duty cycles shorter than $10^{8}$ yr~\citep{Gaspari2017,Prasad2020}, such that periods of low radio activity may fall below survey detection limits. Furthermore, while the highest mass BCGs are frequently radio-active, their luminosity can fluctuate by several orders of magnitudes over these timescales~\citep{Hogan2015,Main2017}.

\subsection{Infrared: Spitzer/WISE}
SPT\,J0417 was observed using the \emph{Spitzer Space Telescope} (PID 70053; PI Brodwin). Infrared photometry was obtained with IRAC~\citep{Fazio2004} in the \(3.6~\mu\)m and \(4.5~\mu\)m bands. Infrared-bright AGN in a cluster at \(z \sim 0.58\) can be identified using a color criterion of \([3.6~\micron] - [4.5~\micron] \geq 0.61\) (Floyd et al., in prep.). Based on this criterion, neither the BCG nor the three ASKAP radio sources exhibit infrared colors consistent with an infrared-bright AGN. However, two additional galaxies in the vicinity exhibit IRAC colors consistent with infrared-bright AGN, located at \(\mathrm{RA} = 04^{\mathrm{h}}17^{\mathrm{m}}22\fs74\), \(\mathrm{Dec} = -47\degr50\arcmin43\farcs0\) and \(\mathrm{RA} = 04^{\mathrm{h}}17^{\mathrm{m}}16\fs79\), \(\mathrm{Dec} = -47\degr50\arcmin46\farcs5\), corresponding to projected cluster-centric distances of 0.743 Mpc and 0.863 Mpc, respectively. Both sources lie outside the field of view presented in the middle panel of Figure~\ref{fig:radio_ir}. An ongoing analysis is underway to assess whether these sources are cluster members or projected interlopers. The broader implications of AGN activity within the cluster environment will be discussed in Floyd et al. (in prep.).

The BCG was also observed with WISE, an all-sky infrared survey with imaging capabilities, at 3.4, 4.6, 12, and 22 \(\rm{\mu m}\)~\citep{Wright2010}. \citet{McDonald2016} estimated the star formation rate (SFR) of the BCG by extrapolating a power-law fit from the observed 22 \(\rm{\mu m}\) data to the rest-frame 24 \(\rm{\mu m}\), following the method of \citet{Calzetti2007}. The non-detection at 22 \(\rm{\mu m}\) sets an upper limit on the SFR of \(28~M_{\odot}~\rm{yr}^{-1}\).

\subsection{Optical: Photometric Modeling}\label{sec:sfr}
\begin{figure*}
\centering
\includegraphics[width=0.95\textwidth]{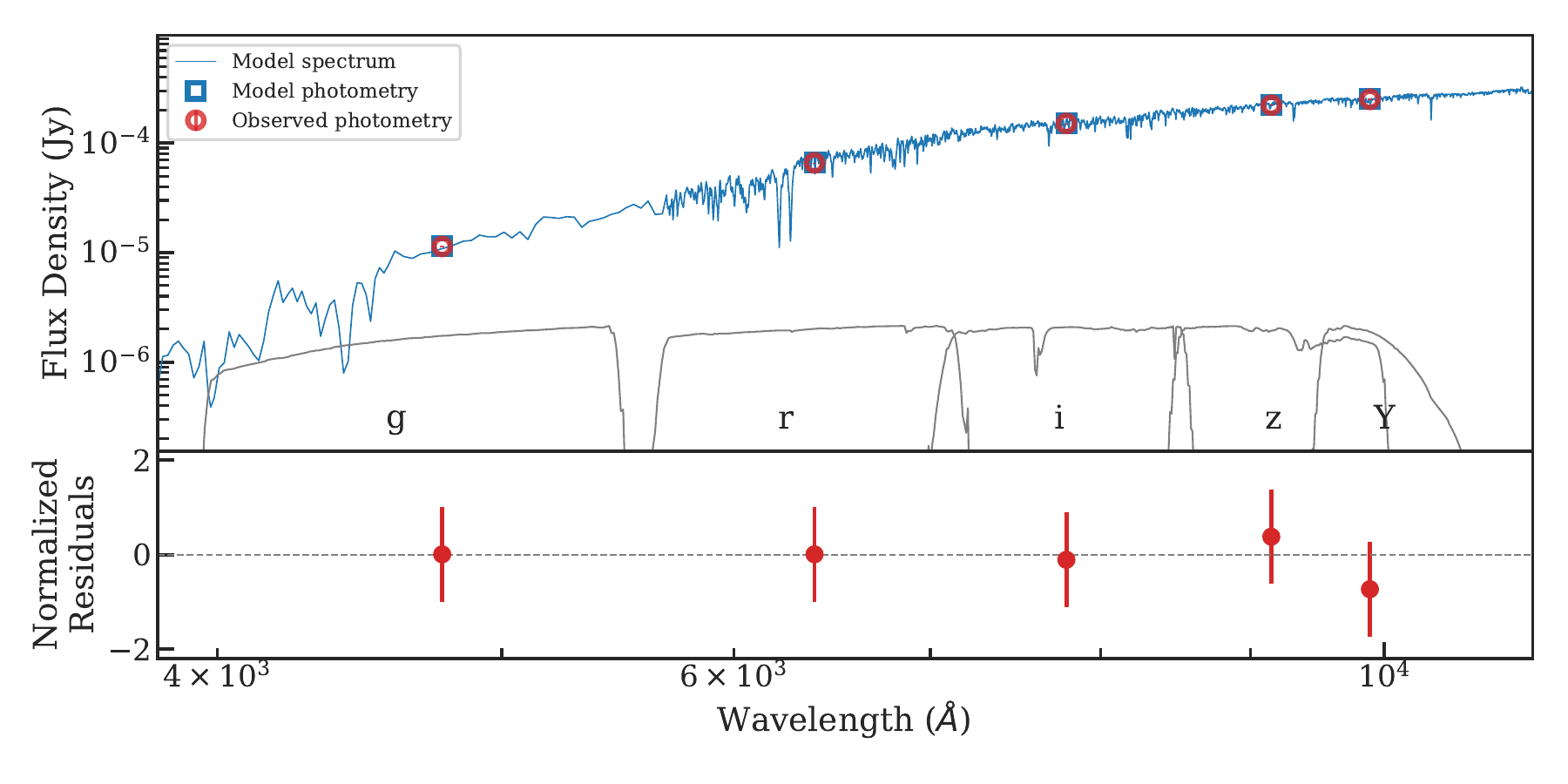}
\caption{Top: Observed-frame optical photometry of the BCG in SPT\,J0417 (DECam $g,r,i,z,Y$; red points with error bars) compared with the best-fit simple stellar population (SSP) model derived using {\sc prospector}. The model spectrum is shown in blue, with the corresponding synthetic photometry indicated by blue squares. The relative transmission curves of each filter are shown in gray for reference. Bottom: Residuals normalized to the photometric uncertainties. The fit is consistent with an old stellar population in the BCG, with little or no evidence of star formation within the last $\sim\!3$ Gyr.
\label{fig:prospector}}
\end{figure*}

We investigated the star formation history of the BCG in SPT\,J0417 by fitting its spectral energy distribution (SED) using $g,r,i,z,Y$ photometry from the Dark Energy Survey DR2~\citep{Abbott2021}. The magnitude was first corrected for Galactic extinction in the direction of the BCG with $A_V=0.0446$~\citep{Schlafly1012.4804}, applying the~\citet{Cardelli1989} extinction curve and $R_V = 3.1$. The modeling was carried out with {\sc prospector} code~\citep{Johnson2012.01426}, adopting a Simple Stellar Population (SSP) template~\citep{Conroy0809.4261,Conroy0911.3151}. In the fits, we assumed a~\citet{Kroupa0009005} initial mass function (IMF) and parameterized internal dust attenuation using~\citet{Kriek2013} model. Posterior distributions of total formed stellar mass, metallicity, and stellar population age were sampled with the {\sc emcee} MCMC package~\citep{Foreman-Mackey1202.3665}. The resulting best-fit SSP spectrum and photometry are shown in Figure~\ref{fig:prospector}. The BCG is found to have a stellar metallicity of $\log_{10}(Z/Z_{\odot}) = -0.10^{+0.26}_{-0.23}$, a total formed stellar mass of $M*=9.5^{+1.2}_{-0.8}\times 10^{11} M_{\odot}$, and a stellar population age of $t_\mathrm{age}=3.0^{+0.8}_{-0.5}$ Gyr, with a reduced chi-squared value of $\chi^2_{\nu}=0.69$. These results suggest that the bulk of the stellar mass in the BCG was assembled around $z=1.25^{+0.33}_{-0.15}$, with little to no subsequent star formation over the past $\sim\!3$ Gyr.
% tage 3.0+0.8-0.5 Gyr => z=0.58 -> t = 7.88     Gyr, meaning that z = (1.097, 1.25, 1.58)

Additionally,~\citet{McDonald2016} obtained an optical spectrum of the BCG in SPT\,J0417 with the Inamori-Magellan Areal Camera and Spectrograph (IMACS) on the 6.5m Magellan Telescopes, targeting the redshifted [O II] emission line as a tracer of star formation in the central galaxy. No [O II] emission was detected, placing an upper limit on the star formation rate of \(<3.8~M_{\odot}~\rm{yr}^{-1}\), consistent with negligible ongoing activity in the BCG. Taken together, these optical results indicate that the BCG in SPT\,J0417 has experienced no significant star formation during the past $\sim\!3$ Gyr.

\section {X-ray Analysis and Results}~\label{sec:results}
The results of our Chandra X-ray analysis are divided into four sections. Section~\ref{sec:spec_analysis} describes the spectral modeling framework adopted in the subsequent analysis. Section~\ref{sec:thermal_profiles} details the thermodynamic structure of SPT\,J0417 through radial profiles of electron density, temperature, and metallicity. Section~\ref{sec:cooling} explores the cooling properties of the intracluster medium, focusing on the pseudo-entropy, cooling time, and free-fall time. These results are based on an analysis that does not assume hydrostatic equilibrium, thereby minimizing potential biases in the temperature profile owing to the surface brightness morphology. Section~\ref{sec:NFW} presents the results derived under the assumption of hydrostatic equilibrium and an NFW mass profile, providing measurements of the total mass, gas mass fraction, and concentration parameter. The characteristic radii $r_{2500}$ and $r_{500}$, within which the overdensity is 2500 and 500 times the critical density of the universe, respectively, were adopted from the best-fitting NFW mass model and used consistently throughout the analysis. We further compare the properties of SPT\,J0417, measured within $r_{500}$, with a sample of 40 dynamically relaxed cool-core clusters from~\citet{Mantz2016} (hereafter \Msixteen{}) via scaling relations.

\subsection{Spectral Analysis and Modeling}~\label{sec:spec_analysis}
For the spectral analysis, we used \texttt{XSPEC} to model the thermal emission from the hot ICM and the local Galactic halo using the \texttt{APEC} plasma model. Photoelectric absorption by Galactic gas was accounted for using the \texttt{XSPEC PHABS} model, with the hydrogen column density fixed to \(N_H=1.28\times10^{20}~\rm{cm}^{-2}\)~\citep{HI4PI1610.06175}. We forward-modeled the three main components of foreground and background emission, namely diffuse Galactic emission, unresolved extragalactic sources and particle-induced backgrounds from cosmic-ray particles, following the methods described by~\citet{Mantz2025}. We adopt the BCG center, located at \(\mathrm{RA} = 04^{\mathrm{h}}17^{\mathrm{m}}23\fs09\), \(\mathrm{Dec} = -47\degr48\arcmin47\farcs6\), as the cluster center for our analysis.

Following the methodology of~\citet{Mantz2014,Mantz2016}, we defined a set of annuli, spaced approximately logarithmically, extending from the cluster center to the radius where the cluster emission represents a 2\(\sigma\) excess over the background. The cluster is modeled as a series of concentric, isothermal, spherical shells with inner and outer radii corresponding to those of the annuli. The expected signal in each annulus was obtained by projecting the 3D model onto the plane of the sky. The density of each shell is constrained individually, whereas the temperature and metallicity are linked between groups of adjacent shells to improve the constraints on these parameters. Beyond \(75^{\prime\prime}\), the cluster signal is no longer significantly detected over the background at most energies; therefore, spectral fitting is not performed, though we do use the integrated 1--2 keV surface brightness to constrain the gas density out to $150^{\prime\prime}$. This region is excluded from the visualization of the temperature and metallicity profiles (and related quantities) because the soft-band surface brightness alone does not provide useful constraints on these parameters in the relevant temperature and redshift regime. 

We conducted two spectral analyses of the cluster: one using the non-parametric \texttt{projct} model in \texttt{XSPEC}, in which the ICM density and temperature profiles are a priori independent, and another in which these profiles are constrained under the assumption of hydrostatic equilibrium with an NFW mass profile, implemented using a modified version of the \texttt{XSPEC} \texttt{nfwmass} model~\citep{Mantz2014}. Following~\citet{Mantz2014}, we exclude the central \(14^{\prime\prime}\) or $\sim$92 kpc region from the NFW-based analysis to mitigate the effects of potential deviations from the hydrostatic equilibrium, spherical symmetry, or the NFW mass model in this high signal-to-noise region. This exclusion radius is determined iteratively by increasing its size until the best-fitting NFW model parameters are stabilized.

% \begin{deluxetable*}{ccc}
% \tablecaption{Central properties of SPT\,J0417 \label{tab:central_properties}}
% \tablehead{
% \colhead{$K_{0}$} & \colhead{$t_{\rm cool,0}$} & \colhead{SFR} \\ 
% \colhead{($\rm keV\,cm^{2}$)} & \colhead{(Gyr)} & \colhead{$(\rm M_\odot yr^{-1})$}
% }
% \startdata
% & & $<3.8$ \\
% \enddata
% \tablecomments{The subscript ``0” denotes the measured quantity in the innermost annulus, with all data deprojected. \(K_{0}\) and \(t_{\rm cool,0}\) represent the central entropy and central cooling time, respectively.}
% \end{deluxetable*}

\subsection{Thermodynamical Profiles, Metallicity} \label{sec:thermal_profiles}

\begin{figure*}[ht!]
\includegraphics[width=0.32\textwidth]{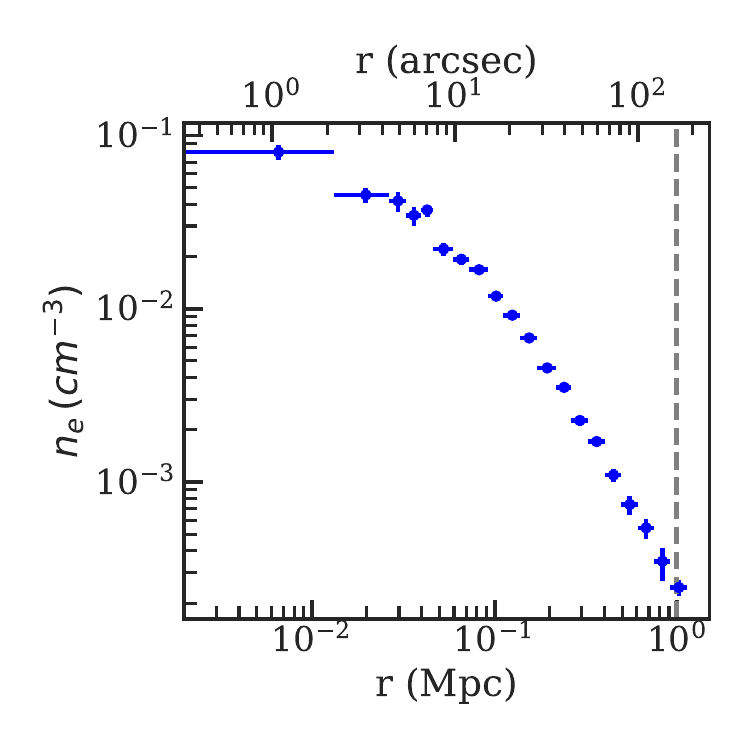}
\includegraphics[width=0.32\textwidth]{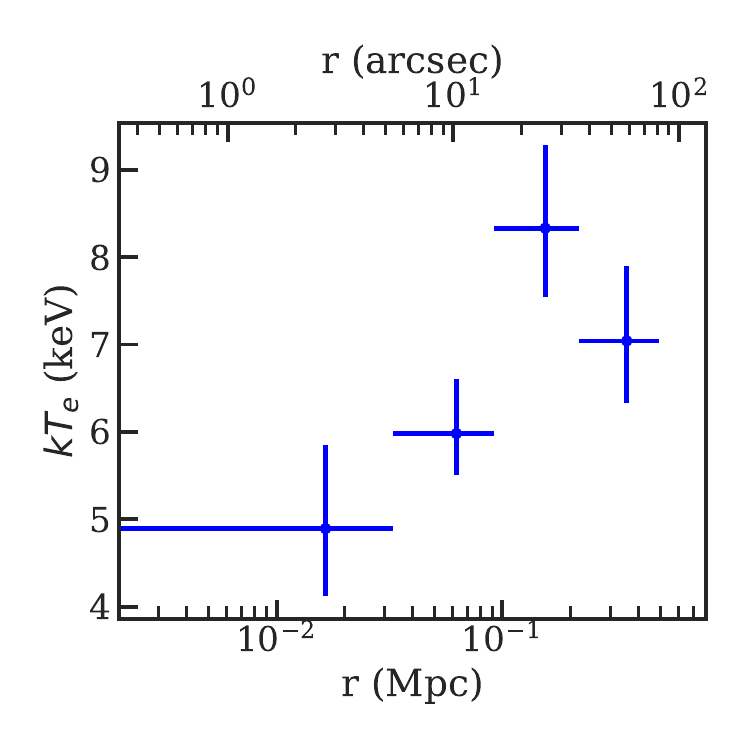}
\includegraphics[width=0.32\textwidth]{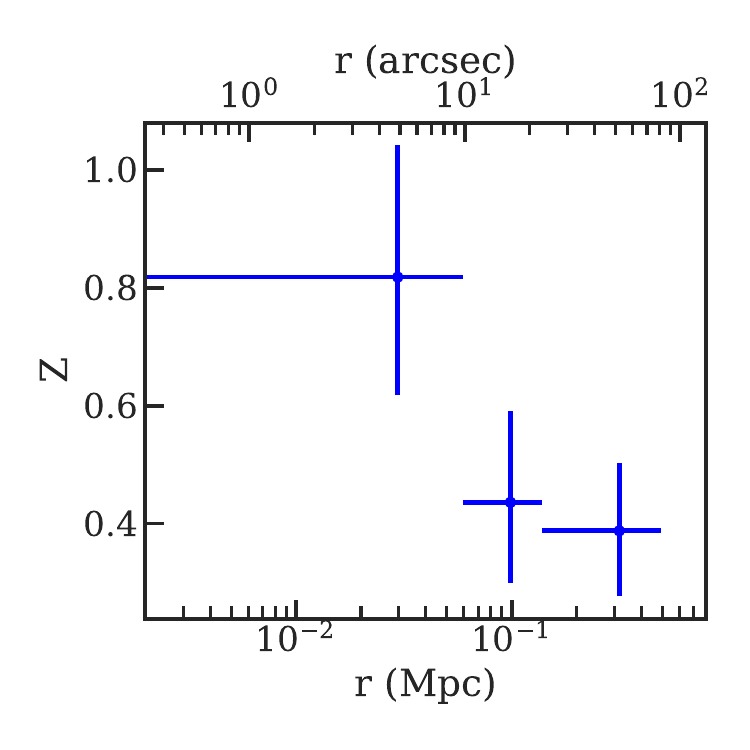}
\caption{Deprojected ICM thermodynamic profiles of SPT\,J0417. Left: Electron density profile, with the gray dashed line marking \( R_{500} \). Middle: Temperature profile showing a clear increase in temperature from the core to the outskirts. Right: Metallicity profile, displaying an enhancement near the cluster core, while the outer regions are consistent with an average metallicity of $\sim\!0.3$ solar~\citep{Mernier2017}.
\label{fig:density_temp}}
\end{figure*}

\begin{figure}[ht!]
\centering
\includegraphics[width=0.45\textwidth]{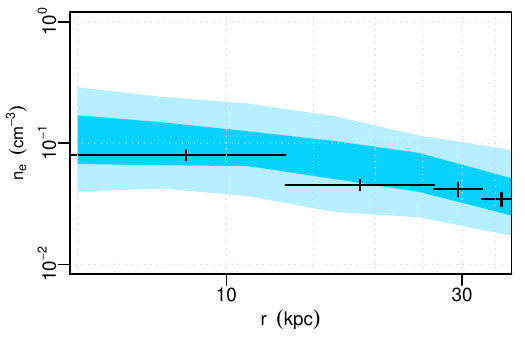}
\caption{Zoom-in view of the central electron density profile of SPT\,J0417 (black), shown as a function of radius, compared with the 68.3$\%$ and 95.4$\%$ confidence ranges of the physical density profiles from the \Msixteen{} sample.~\label{fig:relaxed_zoom}} 
\end{figure}

As shown in the left panel of Figure~\ref{fig:density_temp}, the electron density profile of the cluster features a core in the innermost bin, corresponding to the central \(2^{\prime\prime}\) with a peak density of \(0.08\pm0.01\) \(\rm{cm^{-3}}\). This value is typical when compared to the 40 dynamically relaxed cool-core clusters from \Msixteen{} (see Fig.~\ref{fig:relaxed_zoom}). Owing to the limited number of photons in this region, we combined the first two bins to obtain a single central \(2^{\prime\prime}\) bin with sufficient signal-to-noise. Deeper observations will be necessary to achieve higher spatial resolution and to resolve the structure within the central few arcseconds. 

% We note that our deprojection center used in this analysis, based on the centroid identified via the SPA method, is offset by approximately $0.76^{\prime\prime}$ from the small-scale peak of the X-ray emission. \fix{This choice of centering may introduce a slight apparent flattening in the central density and derived quantities.}

In addition, the middle panel of Figure~\ref{fig:density_temp} presents the temperature profile, which indicates the presence of a strong cool core. The temperature rises steeply from a minimum of $4.8^{+1.0}_{-0.8}$ keV at \(\sim\!20\) kpc to a maximum of $8.3^{+0.9}_{-0.8}$ keV at \(\sim\!200\) kpc. The temperature profile remains well-constrained out to \(0.5r_{500}\), where $r_{500}$ is located at \(\sim\!0.99\) Mpc. 
% with a core-excised (0.15–1.0 \(r_{500}\)) average temperature of \(7.4 \pm 0.5\) keV

Finally, the right panel of Figure~\ref{fig:density_temp} presents the deprojected radial profile of the ICM iron abundance, in solar units~\citep{Asplund0909.0948}. %as measured from the equivalent width of the Fe K$\alpha$ emission line at $\sim\!6.4$\,keV. 
The metallicity profile exhibits a clear radial gradient, decreasing from \(0.8\pm0.2\) solar in the central \(r < 60\) kpc to $0.39\pm0.11$ solar in the outer regions. This outer value is consistent with the average core-excised metallicity observed in lower redshift galaxy clusters ($\sim\!0.3$ solar) and is consistent with the picture in which the accreting gas at large radii is pre-enriched to 0.3 solar, likely through early feedback processes or galactic winds~\citep{,Werner1310.7948,Urban1706.01567}. The elevated central abundance, on the other hand, is likely to reflect additional enrichment from star formation within the central cluster galaxies, as well as from the infalling cores of subhalos that deposit metal-rich gas into the cluster center during merger events.~\citep{Leccardi0806.1445,Ettori1504.02107,McDonald1603.03035,Mernier2017,Mantz2017}.

\subsection{Pseudo-Entropy, Cooling Time, Free-fall Time} \label{sec:cooling}
Several key properties were calculated to investigate the heating and cooling mechanisms within the ICM, including pseudo-entropy, cooling time, and free-fall time. These quantities provide insights into the thermal history and stability of the cluster core.

After obtaining the electron density and temperature profiles, the pseudo-entropy of the ICM is derived using 
\begin{equation}
  K(r) = kT(r) \times n_e(r)^{-2/3},
\end{equation}
\citep{Balogh1999, Ponman9810359} where \( kT(r) \) is the temperature profile, and \( n_e(r) \) is the electron density profile. Pseudo-entropy serves as a key observable for studying the effects of feedback in clusters, as the thermal history of a cluster is governed solely by heat gains and losses~\citep{Cavagnolo2009,Panagoulia2014}. In the absence of feedback mechanisms, the pseudo-entropy profile is expected to increase monotonically with radius~\citep{Voit2005,Cavagnolo2009}. %%==>

\begin{figure*}[ht!]
\includegraphics[width=0.46\textwidth]{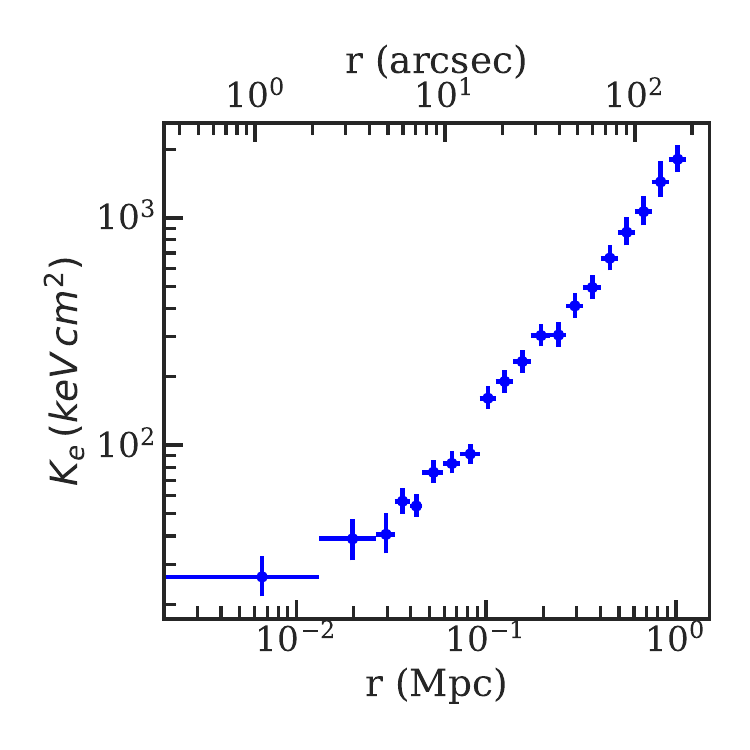}
\includegraphics[width=0.46\textwidth]{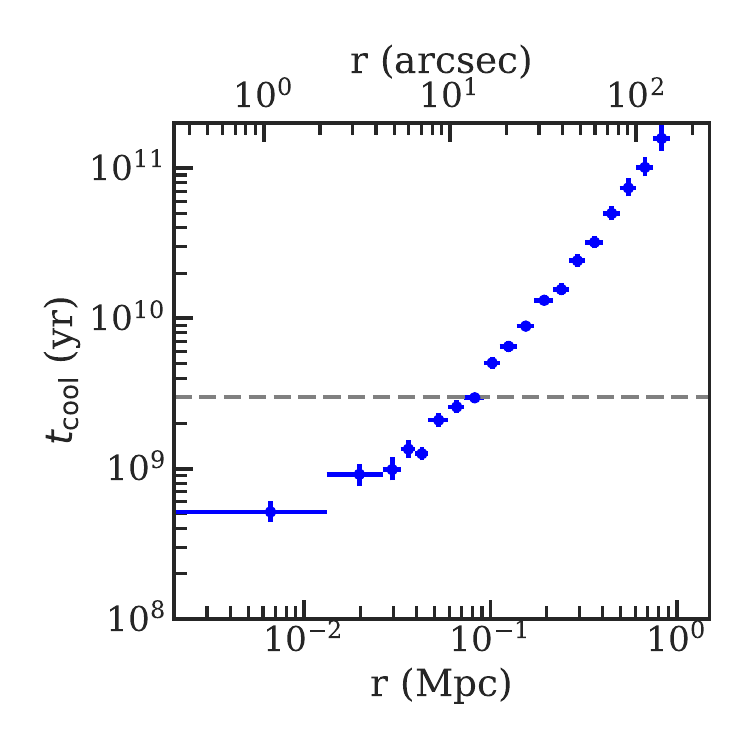}
\caption{Left: Deprojected pseudo-entropy profile of SPT\,J0417. Right: Radial profile of the ICM cooling time. The grey dashed line marks 3 Gyr, which we use to define the cooling radius \(r_{\rm cool}\), as the radius where the cooling time drops below this threshold, for comparison with literature values. The entropy decline to below $\sim\!30\,\rm{keV\,cm^{2}}$ in the core, together with a central cooling time shorter than 1 Gyr, classifies this system as a cool-core cluster~\citep{Hudson2010}. We note that, owing to the limited data, the innermost temperature bin corresponds to $\sim\!30$ kpc. This results in an artificially flat temperature profile in the core, leading to an overestimate of the central pseudo-entropy and cooling time, as the true temperature is expected to decline further toward the cluster center.
\label{fig:entropy_tcool}}
\end{figure*}

The left panel of Figure~\ref{fig:entropy_tcool} shows the radial pseudo-entropy profile derived from the deprojection analysis. Within the central \(r < 13\) kpc, the pseudo-entropy in the innermost bins is measured to be \(26^{+6}_{-5}\,\rm{keV\,cm^{2}}\). At large radii, the pseudo-entropy profile exhibits the expected power-law increase, consistent with predictions from numerical simulations~\citep{Voit2002,Voit0511252} and observations of other clusters (e.g., \citealt{Su1503.03145,Tchernin1606.05657}), for which the ICM entropy profile typically scales as $\propto r^{1.1}$. 
% , which is a typical values found in relaxed cool-core systems, as shown in Figure~\ref{fig:relaxed_scaled}.
%%K_e = 1.26 ± 0.20 * r^1.02 ± 0.03

The cooling time is typically defined in terms of the specific heat in constant volume as
\begin{equation}
t_{\rm cool} = \frac{3}{2}\frac{(n_e + n_p)kT(r)}{n_e n_p \Lambda(T)}, 
\end{equation}
where \( \Lambda(T) \) is the cooling function at temperature \( T \), calculated using the metallicity derived from our spectral analysis~\citep{Sutherland1993}, and \( n_e \) and \( n_p \) are the electron and proton number densities, respectively. The central cooling time, often defined as the cooling time within the innermost $\sim$10 kpc, serves as a key diagnostic tool for distinguishing between cool-core and non-cool-core clusters. As shown in the right panel of Figure~\ref{fig:entropy_tcool}, the central cooling time measured within the innermost annulus (\(r < 13\) kpc) is $515^{+96}_{-75}$ Myr, falling below the 1 Gyr threshold commonly used to classify a system as a cool-core cluster~\citep{Hudson2010}.

An additional quantity of interest is the free-fall time, \( t_{\rm ff} \), which characterizes the timescale for gas to collapse under gravity and is defined as \( t_{\rm ff} = (2r/g)^{1/2} \), where \( g \) is the local gravitational acceleration at radius \( r \). Thermal instability and subsequent cooling of the ICM, leading to the formation of molecular gas and new stars, are observed to occur when the ratio \(t_{\rm cool}/t_{\rm ff}\) drops below a critical value of approximately 10~\citep{Gaspari2012,Sharma2012}. This cooling process can, in turn, fuel feedback from AGN, which reheats the core, suppresses further cooling, and thereby initiates a self-regulating feedback cycle~\citep{Voit2015,Li2015}. This behavior has been reproduced in simulations, which typically yield a minimum \(t_{\rm cool}/t_{\rm ff} > 10\)~\citep{Gaspari2012,Gaspari2013,Li2015}, and is also supported by observations of cool-core clusters exhibiting nebular emission, where the inner \(t_{\rm cool}/t_{\rm ff}\) approaches, but does not fall below, this threshold~\citep{Voit2015,Hogan2017,Pulido2018}. Notably, the Phoenix cluster is the only known exception to this trend, exhibiting \(t_{\rm cool}/t_{\rm ff}\) values approaching unity on kpc scales, while simultaneously hosting substantial star formation in its central galaxy~\citep{McDonald2019}. In the case of SPT\,J0417, we find that \(t_{\rm cool}/t_{\rm ff}\) profile does not fall below the critical threshold of $\sim\!10$ at any radius, which is consistent with the presence of a self-regulated AGN feedback mechanism.

%This suggests that the ratio for SPT\,J0417 remains consistent with the expectation.

To further quantify the cooling properties, we estimated the classically inferred ICM cooling rate using \(\dot{M}_{\rm cool} = M_{\rm gas}(r<r_{\rm cool})/t_{\rm cool},\) where \(r_{\rm cool}\) is the radius at which the cooling time falls below 3 Gyr, corresponding to \(85\pm4\) kpc. Following the convention of~\citet{McDonald2018}, we adopt a cooling time threshold of 3 Gyr to describe the region of the cluster core where cooling is most relevant. Under this definition, we derive a cooling rate of \(560\pm70~M_{\odot}~\rm{yr}^{-1}\). When compared to the upper limit on the star formation rate of 3.8\(~M_{\odot}~\rm{yr}^{-1}\), this implies a cooling efficiency, defined as \(\epsilon_{\rm cool}=\rm{SFR}/\dot{M}_{\rm cool}\), of less than 0.7\%. This value falls within the typical range observed for galaxy clusters (0.3\%--5.8\%;~\citealt{Calzadilla2022}), indicating that SPT\,J0417 is consistent with a quenched system, similar to the majority of the cluster population.

\begin{figure}[htbp]
\centering
\includegraphics[width=0.4\textwidth]{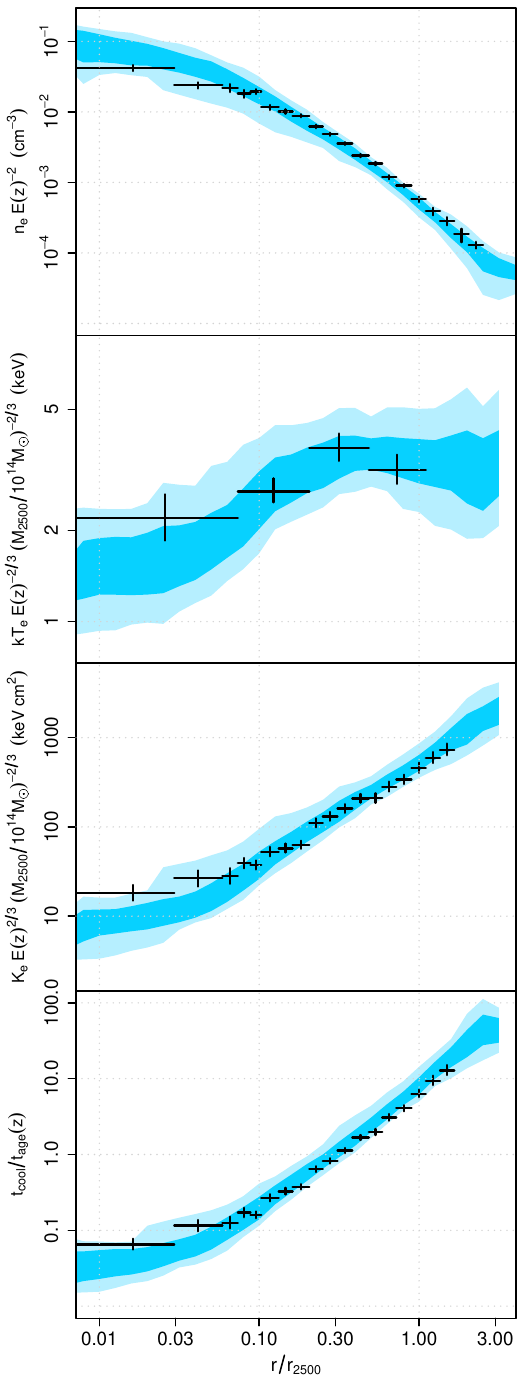}
\caption{Scaled electron density, temperature, pseudo-entropy, and cooling time profiles as a function of scaled radius for SPT\,J0417 (cyan) compared to the 68.3$\%$ and 95.4$\%$ confidence regions of a larger sample of relaxed, cool-core clusters spanning a redshift range of 0.078–1.063 of the \Msixteen{} sample. Relative to this sample, the overall profile shapes of SPT\,J0417 remain consistent with the general trends observed across the larger sample at various redshifts.~\label{fig:relaxed_scaled}} 
\end{figure}

Figure~\ref{fig:relaxed_scaled} compares the scaled thermodynamic profiles of SPT\,J0417 (electron density, temperature, pseudo-entropy and cooling time) with those of 40 other dynamically relaxed, cool-core clusters from~\Msixteen{} (cyan band), spanning a redshift range of 0.078--1.063. Radii are scaled by $r_{2500}$, and the profiles are scaled for the redshift and mass dependence predicted by self similarity. In comparison to this sample, SPT\,J0417 appears broadly consistent, exhibiting the characteristics of a typical dynamically relaxed, cool-core cluster from its inner core out to the outskirts, with central values in good agreement with the comparison set. Note that the innermost temperature bin extends to $\sim\!30 kpc$, which may slightly overestimate the central pseudo-entropy and cooling time (see Figure~\ref{fig:entropy_tcool}).
%Due to the combination of higher redshifts and shallower exposures in parts of that sample, these profiles are typically constrained only to radii larger than $\sim$6 kpc. 

Several diagnostic criteria predict the conditions under which multiphase cooling occurs in ICM. For instance, \citet{Cavagnolo2008} identified either a central pseudo-entropy threshold of 30 \(\rm{keV\,cm^{2}}\), or \(t_{\rm cool} \approx 1\) Gyr, below which strong cooling and significant star formation rates are typically observed. In the case of SPT\,J0417, the core pseudo-entropy and central cooling time approach these thresholds, suggesting the potential for cooling or star formation near the central galaxy of the cluster to occur. However, no evidence for any star formation is detected in the optical spectrum of the central galaxy.

\subsection{Results from Mass Modeling and Scaling Relations} \label{sec:NFW}
%fgas = 0.13+0.01-0.02, c = 4.9+0.8-0.5
In this section, we utilize the \texttt{nfwmass} model within XSPEC, assuming an NFW mass profile and hydrostatic equilibrium, to constrain the characteristic radius, total mass, gas mass fraction, and NFW concentration parameter, defined as \(c = r_{200}/r_s\).

We measure the gas mass fraction, \(f_{\rm gas} = M_{\rm gas}/M_{\rm tot}\), within a spherical shell spanning 0.8–1.2\(r_{2500}\) to be \(0.13^{+0.01}_{-0.02}\), consistent with the measurements of other clusters at similar redshifts~\citep{Mantz2022}. The NFW concentration parameter for SPT\,J0417 was determined to be \(4.9^{+0.8}_{-0.5}\), in agreement with the typical values observed for massive clusters at comparable redshifts~\citep{Darragh-Ford2023}. Together, these results suggest that, on large scales, SPT\,J0417 exhibits properties characteristic of a typical strong cool-core dynamically relaxed cluster, standard gas mass fraction, and concentration consistent with the massive relaxed cluster population.

\begin{figure*}[ht!]
\centering
\includegraphics[width=0.98\textwidth]{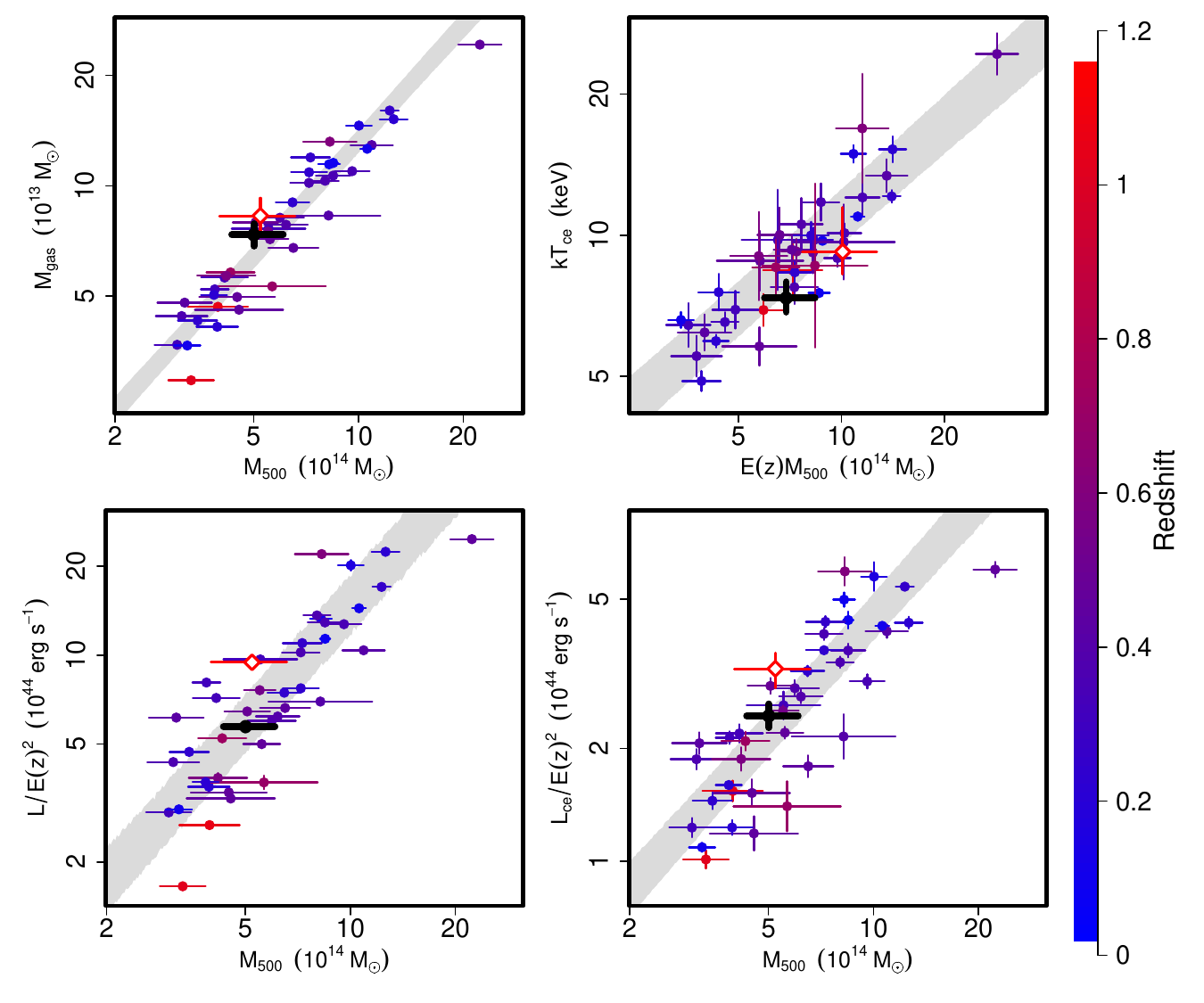}
\caption{Scatter plots of the integrated thermodynamic quantities for SPT\,J0417 (black points) are shown in comparison with the relaxed cluster sample from \Msixteen{}, along with SPT-CL\,J2215-3537 (white-centered red diamond;~\citealt{Stueber2026}). The \Msixteen{} clusters are color-coded by redshift, as indicated by the color bar. The shaded regions represent the 1$\sigma$ predictive intervals derived from a power-law fit to the \Msixteen{} data, accounting for intrinsic log-normal scatter.}
\label{fig:scaling_relaitions}
\end{figure*}

Table~\ref{table:properties} summarizes our measurements of the global properties of SPT\,J0417 derived from the X-ray analysis employing the \texttt{NFWMASS} and \texttt{PROJCT} models, as described. Reported quantities include characteristic masses ($M_{500}$, $M_{200}$), and characteristic radius ($r_{500}$), the core-excised X-ray temperature measured within 0.15--1.0\,$r_{500}$, and the total X-ray luminosity ($L_{X}$) converted to the 0.1–2.4\,keV rest-frame band. To place these measurements in a broader context, we compare them with the relaxed cluster sample of \Msixteen{}, and recent results for SPT-CL\,J2215-3537 (in red;~\citealt{Stueber2026}), in Figure~\ref{fig:scaling_relaitions}. This figure presents scaling relations between the characteristic mass, gas mass, average temperature, total luminosity, and core-excised luminosity within $r_{500}$, with temperatures and core-excised luminosities excluding the central $<0.15,r_{500}$. Clusters from the \Msixteen{} sample are color-coded by redshift, whereas SPT\,J0417 is marked in black. The gray shaded regions indicate the $1\sigma$ predictive intervals from a power-law fit to the \Msixteen{} sample, incorporating intrinsic log-normal scatter. Across all properties, SPT\,J0417 lies within the expected range for massive, relaxed, cool-core clusters over a redshift span of 0.078--1.063~\citep{Mantz2016}.

\begin{deluxetable*}{ccccccc}
\tablecaption{Global properties of SPT\,J0417 \label{tab:cluster_properties}}
\tablehead{
\colhead{$z$} & \colhead{$r_{500}$} & \colhead{$M_{500}$} & \colhead{$M_{200}$} & \colhead{$M_{\rm gas}$} & \colhead{$kT$} & \colhead{$L_X$ (0.1-2.4 keV)}\\ 
\colhead{} & \colhead{(Mpc)} & \colhead{($10^{14}\,M_\odot$)} & \colhead{($10^{14}\,M_\odot$)} & \colhead{($10^{13}\,M_\odot$)} & \colhead{(keV)} & \colhead{($10^{44}$ erg s$^{-1}$)}
}
\startdata
0.581 & $0.99^{+0.05}_{-0.06}$ & $5.0^{+1.1}_{-0.7}$ & $6.9^{+1.6}_{-1.1}$ & $7.4^{+0.5}_{-0.5}$ & $7.4^{+0.6}_{-0.5}$ & $10.8\pm0.3$ \\
\enddata
\tablecomments{Where applicable, measurements are referenced to the characteristic radius \( r_{500} \). The X-ray temperature (\( kT \)) is extracted from the region excluding the cluster core, specifically from 0.15–1 \( r_{500} \).}
\label{table:properties}
\end{deluxetable*}

\section{Conclusion}~\label{sec:conclusion}
In this study, we present new deep \(\sim\!100\) ks observations of SPT\,J0417 using the \Chandra{} X-ray Observatory. Our key findings are as follows: 

\begin{enumerate}
    \item The large-scale X-ray morphology of SPT\,J0417 is consistent with a dynamically relaxed system, as indicated by the SPA criteria~\citep{Mantz2015}. Its global properties, including the total X-ray luminosity ($L_X$), total mass ($M_{500}$), gas mass fraction, and concentration parameter, are broadly consistent with those of typical massive, relaxed, cool-core clusters.
    
    \item The peak ICM density in the core is measured to be \(0.08\pm0.01\) \(\rm{cm^{-3}}\), also typical of cool-core clusters. The central pseudo-entropy (\(26^{+6}_{-5}\,\rm{keV\,cm^{2}}\)) and central cooling time ($515^{+96}_{-75}$ Myr) fall below the thresholds of $\sim\!30$ \(\rm{keV\,cm^{2}}\) and $\sim\!1$ Gyr, respectively, suggesting strong cooling in the core and conditions that could support significant star formation.
    
    \item Both SED fitting of the BCG photometry and the non-detection of [O II] emission in its optical spectrum indicate an absence of recent or ongoing star formation, with the majority of the stellar population formed at $z\sim1.25$ and no significant activity over the past $\sim\!3$ Gyr. These results suggest that cooling in the cluster core has been suppressed to levels insufficient to sustain detectable star formation. In addition, ASKAP radio and Spitzer infrared data show no evidence for strong ongoing AGN activity in the BCG.
    %the absence of a radio detection at the BCG location, combined with the low probability of the BCG being an IR-AGN and , indicate little to no black hole accretion and . 
\end{enumerate}

SPT\,J0417 represents another example of a massive dynamically relaxed, cool-core cluster with a tight upper limit on ongoing or recent star formation in its central galaxy. Comparable systems include Abell 2029, a massive relaxed galaxy cluster at $z = 0.07795$ with only minimal star formation (0.03--0.06 $M_{\odot}~\rm{yr}^{-1}$), as measured with GALEX~\citep{Hicks2010}, and MS2137-2353 at $z = 0.313$, which hosts a BCG with a star formation rate of $\sim\!2\,M_{\odot}~\rm{yr}^{-1}$~\citep{Cooke1610.05310}. SPT\,J0417 extends this population to higher redshift, while maintaining a comparable mass and similar central thermodynamic properties. Such systems stand in contrast to scenarios predicting that the most massive cool-core clusters should exhibit high star formation rates (\(>\!\!100~M_{\odot}~\rm{yr}^{-1})\) due to saturation of AGN feedback heating.
%M500 A2029 = 7.2e14 Msun, 

One possible explanation for the lack of star formation in SPT\,J0417 is hot-mode accretion. In this scenario, the AGN is fueled not by cool gas but by the surrounding hot ICM. While accretion from hot gas alone is generally insufficient to balance ICM cooling, a sufficiently massive central supermassive black hole, with a mass $\sim 10^{10}\,M_{\odot}$, could produce feedback powerful enough to suppress cooling and prevent cold gas formation. Further studies of BCG properties, particularly SMBH masses in relaxed clusters with and without ongoing star formation, will be critical to test this scenario, with SPT\,J0417 providing an important intermediate-redshift case study of this class of systems. 

% \fix{Finally, deeper X-ray observations will be essential for characterizing the slight asymmetry observed in the cluster core and assess whether it corresponds to a weak cold front. More sensitive radio data are also required to probe faint or emission from past AGN activity.}

Finally, deeper and higher spatial resolution radio and deeper X-ray observations of the inner few arcseconds of the cluster are needed to probe AGN activity in the BCG and to search for subtle ICM features such as sloshing or cold fronts in the core. In addition, complementary rest-frame optical and near-infrared spectroscopy with facilities such as the Hubble Space Telescope and the James Webb Space Telescope will also be essential for constraining any residual star formation in the BCG, offering new insight into the processes that regulate cooling and feedback in massive clusters.

\section{Acknowledgments}
TS acknowledges support from the Kavli Postdoctoral Fellowship at Stanford University and from the grants for the development of new faculty staff, Ratchadapiseksomphot Fund, Chulalongkorn University.

This research has made use of data obtained from the \Chandra{} Data Archive provided by the Chandra X-ray Center (CXC). Support for this work was provided by the National Aeronautics and Space Administration through \Chandra{} Award Number GO3-24113X issued by the Chandra X-ray Center, which is operated by the Smithsonian Astrophysical Observatory on behalf of the National Aeronautics Space Administration under contract NAS8-03060. This paper employs a list of \Chandra{} datasets, obtained by the \Chandra{} X-ray Observatory, contained in the \Chandra{} Data Collection (CDC) `378'~\dataset[doi:10.25574/cdc.378]{https://doi.org/10.25574/cdc.378}. We acknowledge support from the U.S. Department of Energy under contract number DE-AC02-76SF00515.

This work was performed in the context of the South Pole Telescope scientific program. The South Pole Telescope program is supported by the National Science Foundation (NSF) through awards OPP-1852617 and OPP-2332483. Partial support is also provided by the NSF Physics Frontier Center grant PHY-0114422 to the Kavli Institute of Cosmological Physics at the University of Chicago, the Kavli Foundation and the Gordon and Betty Moore Foundation grant GBMF 947 to the University of Chicago. The SPT is also supported by the U.S. Department of Energy. Argonne National Laboratory’s work was supported by the U.S. Department of Energy, Office of High Energy Physics, under contract DE-AC02-06CH11357.
% \end{acknowledgments}

%% in the research for the paper.Each keyword is check against the master 
%% list during copy editing.Individual instruments can be provided in 

% \vspace{5mm}
\facilities{CXO, Spitzer, WISE, ASKAP, Magellan 6.5m telescopes (Baade/IMACS)}

%% Similar to \facility{}, there is the optional \software command to allow authors a place to specify which programs were used during the creation of the manuscript. Authors should list each code and include either a
\software{
    astropy~\citep{Astropy-Collaboration1307.6212,Astropy-Collaboration1801.02634,Astropy-Collaboration2206.14220}, 
    CIAO~\citep{ciao1311.006, Fruscione2006},
    MARX \citep{marx1302.001, Davis2012SPIE.8443E..1AD},
    LMC \citep{lmc1706.005},
    HEASOFT \citep{heasoft1408.004},
    SXRBG \citep{sxrbg1904.001},
    XSPEC~\citep{xspec9910.005, Arnaud1996}
}

% \appendix

% \begin{figure}[ht!]
% \centering
% \includegraphics[width=0.5\textwidth]{figures/Rplots_spt0417.pdf}
% \includegraphics[width=0.5\textwidth]{figures/Rplots_density_scaled.pdf}
% \caption{
% \label{fig:density_other}}
% \end{figure}


%% the bibliography. The sample631.bib file was populated by ADS. To

\bibliography{sample631}{}
\bibliographystyle{aasjournal}

%% the collaboration and author truncation commands are used.It has to


\end{document}